\documentclass[aps,pre,twocolumn,showpacs,superscriptaddress]{revtex4}

\usepackage{graphicx}
\usepackage{amssymb}

\newcommand{\figwidth}{0.9\linewidth}

\begin{document}

\title{Shortest paths and load scaling in scale-free trees}

\author{G\'abor Szab\'o}
\affiliation{Helsinki University of Technology, Laboratory of Physics,
P.O.Box 1100, FIN-02015 HUT, Finland}
\affiliation{Budapest University of Technology, Department of
Theoretical Physics, H-1111 Hungary}

\author{Mikko Alava}
\affiliation{Helsinki University of Technology, Laboratory of Physics,
P.O.Box 1100, FIN-02015 HUT, Finland}

\author{J\'anos Kert\'esz}
\affiliation{Budapest University of Technology, Department of
Theoretical Physics, H-1111 Hungary}

\date{\today}

\begin{abstract}
\noindent
The average node-to-node distance of scale-free graphs depends
logarithmically on $N$, the number of nodes, while the probability
distribution function (pdf) of the distances may take various
forms. Here we analyze these by considering mean-field arguments and
by mapping the $m=1$ case of the Barab\'asi-Albert model into a tree
with a depth-dependent branching ratio. This shows the origins of the
average distance scaling and allows a demonstration of why the
distribution approaches a Gaussian in the limit of $N$ large. The {\em
load}, the number of shortest distance paths passing through any node,
is discussed in the tree presentation.
\end{abstract}

\maketitle

\section{Introduction}

Recently, many examples have been found of systems whose innate
topology is not homogenous and can rather be described in terms of a
scale-free, random structure. Examples range from the Internet to
cellular metabolism networks. The interest of the physics community in
this field stems from the fact that the behavior of many systems on
such networks or {\em graphs} changes drastically and often attains
characteristics close-to but not quite like the mean-field limit.

A scale-free graph consists of a set of nodes or vertices $V$ and
bonds or edges $E$ connecting the vertices to a structure. The
essential measure of the scales or lack thereof is the connectivity or
degree distribution of the nodes $V$: the probability of any node to
have $k$ edges (one may distinguish between directed and undirected
graphs; in the former case the in-going and out-going pdf's can
differ). If this probability $P_k$ follows a power-law behavior, a
structure arises that does not have any intrinsic scale. The Internet
is an example of such a $P_k \sim k^{-\gamma}$, and several models
have been designed that fit the same description. Later on enhanced
models have been devised to capture the characteristics of more
elaborate phenomena, like the tendency of
clustering~\cite{tunable_clustering,clustered_scale_free,develop_decay,heritable_connectivity}.

The models lead to evolving graphs which grow continuously in time by
the addition of new nodes, with only a limited number of notable
exceptions where the scale-free graph is generated by means of a
Monte-Carlo algorithm~\cite{Krzywicki1}. The degree distribution $P_k$
and average connectivity become stationary in the thermodynamic limit,
save for the tail of the distribution which is subject to finite-size
cutoff effects~\cite{generic_scale,scaling_evolving}. A practically
minded question in the same spirit is the growth mechanism of the
Internet~\cite{internet}.

It is a common feature of growing networks that they spontaneously
develop degree-degree correlations between adjacent
nodes~\cite{stat_mech,evolution}. This is a manifestation of the {\em
preferential attachment}~\cite{BA_original} principle, where more
connected nodes are to attract a larger proportion of new links as the
network grows. One recent study hints that the correlation between
neighboring node connectivities is the mechanism behind the
logarithmic scaling of the {\em network diameter} or the average
shortest distance between two randomly chosen vertices, with respect
to the system size~\cite{Krzywicki2}. The support for the argument is
empirical evidence from simulation results of a broader class of
scale-free graph ensembles, where a power law growth of the diameter
has been indeed identified. The question of viability of logarithmic
scaling in real-world networks is particularly essential, since it has
an impact on efficiency and percolation issues (communication over the
Internet, spreading
phenomena~\cite{epidemic_spreading,epidemic_dynamics}, community
structures~\cite{community_structure,local_events}).

Until recently, less attention has been paid to the {\em probability
distribution} of shortest path lengths or sometimes referred to as
chemical distances in scale-free graphs, possibly owing to the fact
that it has been implicitly assumed that the average diameter is an
adequate measure of distance properties in the networks. The
particular form of the distribution function may have bearings on the
performance of search algorithms in scale-free
graphs~\cite{path_finding}. On the other hand, the distribution of
shortest paths has been analytically calculated for the small-world
model, employing the underlying lattice
structure~\cite{exact_small_world} and arriving at a Gaussian-like
distribution for large system sizes. Likewise, a model for
deterministic scale-free graphs has been proposed and analyzed
lately~\cite{pseudofractal}, where a Gaussian is again obtained in the
asymptotic limit.

In this paper we focus on a subset of scale-free graphs described by
the Barab\'asi-Albert model~\cite{BA_original} that in addition are
{\em loopless rooted trees} from a topological point of view, i.e. the
$m=1$ case where one connects new nodes by only one link to the
existing structure. By removing the redundancy of interconnecting
loops it is possible to consider the distance properties on a
mean-field level, and also to analyze ``load'' or
``betweenness''~\cite{universal_load,scientific_coll}, the number of
any shortest paths passing through vertices. The essential fact here
is that the hub of the tree, e.g. the node with the highest
connectivity for simplicity, transmits connections between all the
branches emanating from it. We show that a stochastic branching
process rooted in the preferential attachment rule gives rise to the
logarithmic scaling of the diameter and that the pdf of the minimal
paths approaches a Gaussian.

Since throughout this text we are interested in tree structures, it is
useful to overview their basic features. In the context of random
networks, one often resorts to Ref.~\cite{arbitrary_degree} and the
derivation therein, which suggests that the diameter of graphs grows
logarithmically. Although the calculations there are performed for
random graphs containing loops, the result obtained closely resembles
that for {\em balanced Cayley trees} with uniform coordination numbers
(except for the coordination number of the central node, which is
different). According to this, the number of nodes separated from node
zero by $k$ nonrecurring steps goes as $z^k$ where $z$ is the
coordination number for the Cayley tree. It then follows simply from
the sum of a geometric series that both the longest distance between
nodes and the average distance should behave as $\bar{l} \sim \ln N$.

It is obvious that trees have unique shortest paths between any two
nodes in the sense that without traversing the same edge twice it is
not possible to find an alternate minimal route (unlike in unweighted
graphs with loops, where there is usually more than one minimal
path). We can then define one of the nodes as the {\em root} of the
tree and unambiguously arrange all the other nodes into {\em layers}
depending on their minimal distance to the root. Finding the shortest
path between two chosen nodes is nothing but identifying the deepest
common node along the paths leading from the root to the source and
target vertices and then connecting the two nodes via this common
fork. Notice that the choice of the root here is slightly arbitrary;
one would prefer to use balanced trees.

We study scale-free Barab\'asi-Albert trees~\cite{BA_original},
starting with a single vertex. In each time step then we add a new
vertex with {\em only one} outgoing edge. The other end of the edge is
connected to one of the nodes already present in the system with a
connection probability proportional to the connectivity or degree of a
particular node. All edges are thought of as bidirectional and having
the same weight, namely 1. As a slight modification to the original
model, the connection symmetry of the first two nodes is broken by
introducing a ``virtual'' edge to the very first vertex which only
gives preference for this node over the second one when it comes to
the subsequent addition of further nodes. This way, we can
automatically identify the most connected node in the network and call
it its root. To have a balanced tree one needs every subtree of the
root to have the same number of nodes in the configurational
average. This is only attained when the root is the most connected
node, since the BA model ensures that the order of the nodes in terms
of connectivity does not change in the course of addition of new nodes
and is fully determined by the time of introduction of a node.

\section{Mean-field approach}

To begin with, we will investigate the shortest path distribution in a
mean-field model of a tree network, between the root of the tree and
all the other nodes. This argument extends also to {\em general
graphs} in the case that the new nodes added (e.g. $m>1$
Barab\'asi-Albert networks) do not cause a significant amount of
shortcuts between already existing nodes.

Let us consider a uniform branching process for each of the layers in
the tree so that every node on a certain layer has the same number of
offsprings to produce the next layer beneath; it shall amount to
$b(l)$ for layer $l$ for short. This way the original stochastic model
is approximated by a deterministic graph~\cite{pseudofractal,BRV}.
The number of nodes $n(l)$ with a separation $l$ from the root is then
$n(l) = n(l-1) \, b(l-1)$ with the condition that $n(0) = 1$. The
actual form of $b(l)$ can be obtained by making use of the
preferential attachment rule for BA networks. According to this, the
probability that a newly introduced node will connect to any given set
of nodes is proportional to the cumulative connectivity of the set in
question. Thus, the number of nodes on layer $l+1$ changes according
to the following rate equation, due to the addition of a new node:
\begin{eqnarray}
\frac{\partial n(l+1)}{\partial N} = \frac{1}{2N} (b(l)+1) \, n(l),
\end{eqnarray}
since the right hand side describes the attachment probability to
layer $l$, where $N$ is the number of nodes in the system and $2N$ is
the normalization factor for the connectivity. Writing $n(l+1) = n(l)
\, b(l)$, expanding the derivation, and dividing by $n(l) \, b(l)$
give
\begin{eqnarray}
\frac{1}{b(l)} \frac{\partial b(l)}{\partial N} =
\frac{1}{2N} \left( \frac{1}{b(l)} - \frac{1}{b(l-1)} \right).
\end{eqnarray}

If we substitute $B(N,l) = 1 / b(l)$ by explicitly indicating the size
dependence on $N$ and assume that $B(N,l)$ is a slowly varying
function,
\begin{eqnarray}
-\frac{\partial \ln B(N,l)}{\partial N} \approx \frac{1}{2N}
\frac{\partial B(N,l)}{\partial l}.
\end{eqnarray}

It is straightforward to expect a solution in the decomposed form of
$B(N,l) = B_l(l) / B_N(N)$:
\begin{eqnarray}
\frac{\partial \ln B_N(N)}{\partial N} \approx \frac{1}{2N}
\frac{1}{B_N(N)} \frac{\partial B_l(l)}{\partial l},
\end{eqnarray}
and since the left hand side is a function of only $N$, $B_l(l)=2
\alpha l$ with a constant $\alpha$. Finally, we get $B_N(N)=\alpha \ln
N$ and
\begin{eqnarray}
b(l) = \frac{1}{2} \frac{\ln N}{l}.
\label{MF_branching}
\end{eqnarray}

This relation does not apply to the root ($l=0$) for obvious
reasons. Eq.~(\ref{MF_branching}) also implies that the number of
nodes with a given distance to the root $n(l)$ keeps growing with $l$
until $b(l)$ drops below $1$ and then starts to decrease, as the
bottom of the tree is approached ($n \ll 1$). This is in strong
contrast to the formal prediction of a constant branching for {\em
any} random graph~\cite{arbitrary_degree}, which would result in a
monotonous exponential growth, as would be the case in usual Cayley
trees.

\section{Distance scaling}

Using the recursion relation $n(l) = n(l-1) \, b(l-1)$ for the number
of nodes on a given level, we can give an estimate for the shortest
path distribution function with the source of the paths at the root of
the tree. Instead of Eq.~(\ref{MF_branching}) we take now the more
general form of $b(l) = (A / l) ^ \lambda$ and approximate the sum
with an integral in the following expression, which reads
\begin{eqnarray}
\lefteqn{n(l) = n(0) \prod_{i=0}^{l-1} b(i) = b(0) \exp \left[
\ln \sum_{i=1}^{l-1} b(i) \right] \approx {}}
\nonumber \\
& & {} \approx b(0) A^{\lambda (l-1)} \exp \left( -\lambda \int_{1}^{l-1}
\ln x \, dx \right) = {}
\nonumber \\
& & {} = \frac{b(0)}{e^\lambda} \left( \frac{A e}{l-1} \right) ^{\lambda
(l-1)}. {}
\label{MF_n_l}
\end{eqnarray}

\begin{figure}
\begin{minipage}[t]{\linewidth}
\begin{center}
\includegraphics[width=\figwidth]{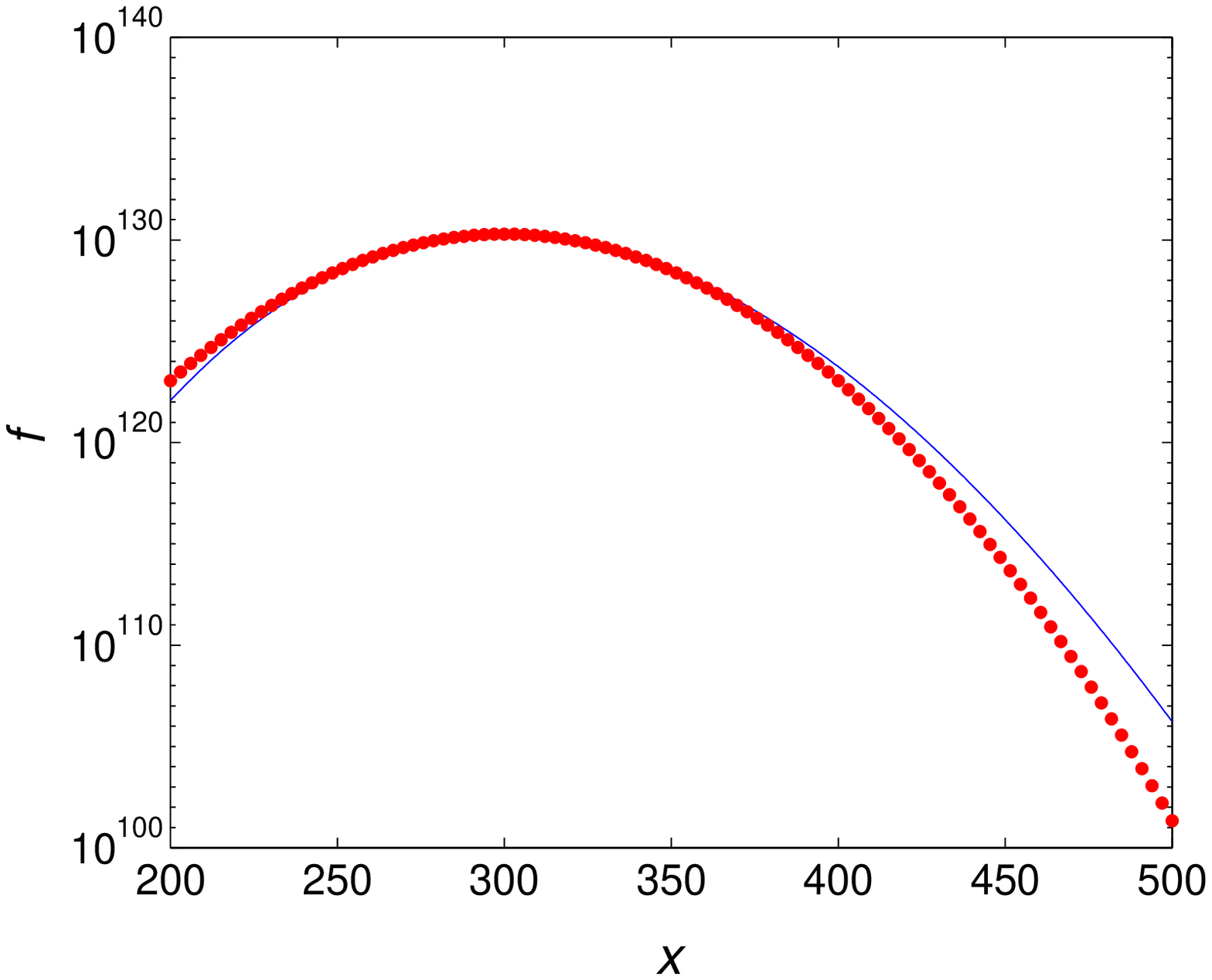}
\end{center}
\vspace{-43mm}
\hspace{-17mm}
\includegraphics[width=0.38\linewidth]{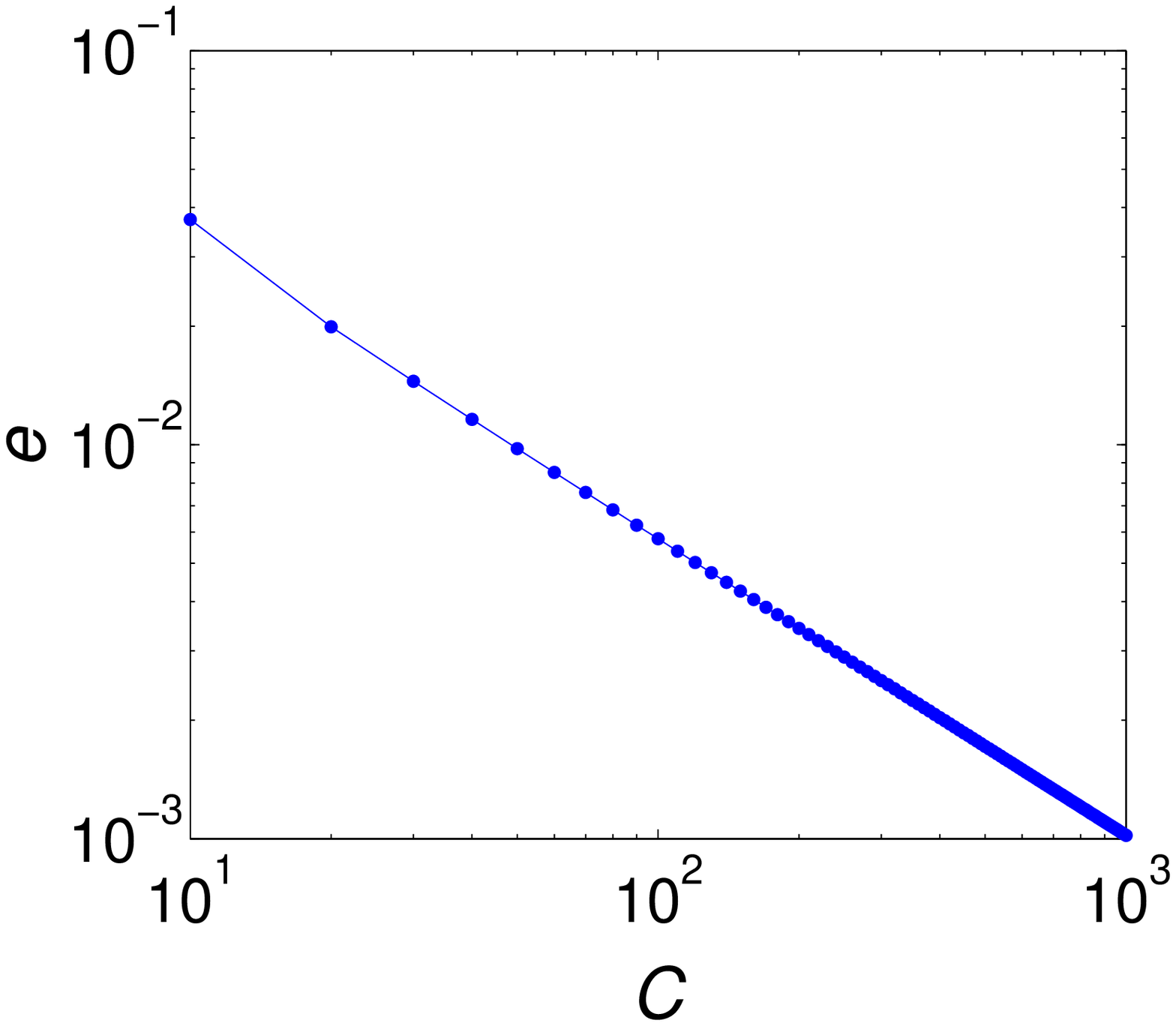}
\end{minipage}
\caption{A Gaussian fit for the function $f(x)=(C/x)^x$ with $C=300 \,
e$. A few points of the Gaussian are represented by the dots. The
difference is only noticeable at the tails of the functions. The inset
shows how the quadratic error of the two functions (normalized for
area) appears to be a decreasing power law with increasing $C$.}
\label{nl_gauss_match}
\end{figure}

The result above for $n(l)$ approaches a non-normalized Gaussian in
the large-$N$ limit as $A \sim \ln N$ goes to infinity, which can be
seen from Fig.~\ref{nl_gauss_match} where a $C$ corresponding to a
very large network has been used. In order to draw further
conclusions, we will determine the parameters of the Gaussian that
give a best fit to $n(l)$. For the sake of simplicity, let us now
consider the function of the form $f(x) = (C / x) ^ {\lambda x}$:
\begin{eqnarray}
f(x) = \left( \frac{C}{x} \right) ^ {\lambda x} \approx
R \exp \left[ -\frac{(x - \mu) ^ 2}{2 \sigma ^ 2} \right].
\label{nl_approx}
\end{eqnarray}

We first match the extremal point of $f$ to the mean of the Gaussian,
resulting in $\mu = C / e$. The maximum value is thus $R=\exp(C
\lambda / e)$; the standard deviation $\sigma$ can be obtained by the
requirement that the derivative functions of $f$ and the Gaussian be
the same in the vicinity of $\mu$ up to first order, giving $\sigma =
\sqrt {C/(\lambda e)}$. Using the parameters acquired this way, we can
find a very good approximation to $f(x)$, which is almost identical to
that of a least-square fit.

Furthermore, additional information can be gained if we look into the
normalization conditions for $n(l)$. Trivially, the sum of $n(l)$ over
all layers should return the total number of nodes in the system,
$N$. Again, we approximate the sum with an integral:
\begin{eqnarray}
N & = & \sum_{l=1}^{\infty} n(l) \approx \int_{1}^{\infty}
n(x) dx = \frac{b(0)}{e ^ \lambda} \int_{0}^{\infty}
\left( \frac{Ae}{x} \right) ^ {\lambda x} \, dx \approx
\nonumber \\
& \approx & \frac{b(0)}{e ^ \lambda} \int_{0}^{\infty} e^{A \lambda}
\exp \left[ -\frac{(x-A)^2}{2 A / \lambda} \right] \, dx \approx {}
\nonumber\\
{} & \approx & \frac{b(0)}{e ^ \lambda} e^{A \lambda}
\sqrt{\frac{2 \pi A}{\lambda}},
\label{N_nl_expr}
\end{eqnarray}
where we assumed that $A$ is large enough so that we can neglect the
correction of the error function to the Gaussian integral. We should
also be aware that $l$ has a finite cutoff because of the bounded
depth of the tree---yet, the quickly vanishing $n(l)$ makes it
possible to take $l$ to infinity. Finally, $f(x) \rightarrow 1$ as $x
\rightarrow 0$ so that the integrand is bounded everywhere.

Recall now that the degree of a node in BA networks grows with the
power of $N$, $b(0) \sim N^\beta$~\cite{stat_mech}. Apart from $b(0)$,
the only term on the right hand side of Eq.~(\ref{N_nl_expr}) that may
contribute to the overall linear growth in $N$ is $e^{A \lambda}$,
which increases much faster than $\sqrt A$, so the latter can be taken
a constant. The consistency condition with the left hand side requires
that $e^{A \lambda} \sim N^{1-\beta}$ should hold, and thus
\begin{eqnarray}
A = \frac{1-\beta}{\lambda} \ln N + \textit{const}.
\end{eqnarray}

Disregarding the constant, we end up with a very similar but more
general expression than that of Eq.~(\ref{MF_branching}) for $b(l)$:
\begin{eqnarray}
b(l) = \left( \frac{\upsilon \ln N}{l} \right) ^ \lambda, \qquad
\textrm{with} \quad \beta + \upsilon \lambda = 1.
\label{exp_relation}
\end{eqnarray}

This implies that if a scale-free tree is characterized by a branching
process decaying as a power law as a function of the distance from a
suitable root node with the highest connectivity, the
relation~(\ref{exp_relation}) should necessarily be satisfied. Not
surprisingly, it is true in the case of BA trees, where $\beta = 1/2$
and according to Eq.~(\ref{MF_branching}), $\upsilon = 1/2$ and
$\lambda = 1$. One should note that in the process of constructing the
mapping we rely on the fact that the number of nodes in a layer
depends only on the {\em average branching ratio} $b(l)$; the
fluctuations in the degrees of nodes is omitted. For this reason the
degree distribution exponent $-3$ is not present in the tree
representation.

The node-to-node distances in the mean-field model are calculated as
follows. We traverse each node of the tree and enumerate the routes
with certain lengths that start at or go through this node and have
both of their ends in the subtree of the node. Practically speaking,
we can think of this node as the root for its subtree and perform the
same calculations as we would do for the 'global root' of the tree. If
we by $n^{(s)}(l)$ denote the number of possible paths going out to
the subtree of a node on level $s$ that have length $l$ and one end
fixed at the node on level $s$,
\begin{eqnarray}
n^{(s)}(l) = \prod_{i=0}^{l-1} b(s+i).
\label{subtree_leaves}
\end{eqnarray}

Let now $r^{(s)}(l)$ be the number of all routes that go {\em through}
or end at a particular node on level $s$ and have a length of $l$,
\begin{eqnarray}
\lefteqn{r^{(s)}(l) = {}}
\nonumber\\
&& {} = n^{(s)}(l) + \Theta(b(s)-1) \sum_{i=1}^{l-1}
{b(s) \choose 2} \frac{n^{(s)}(i)}{b(s)} \frac{n^{(s)}(l-i)}{b(s)} = {}
\nonumber\\
\lefteqn{{} = n^{(s)}(l) + \Theta(b(s)-1)
\frac{b(s) - 1}{2 b(s)}
\sum_{i=1}^{l-1} n^{(s)}(i) \, n^{(s)}(l-i).}
\end{eqnarray}
The second term in the sum has contribution to $r^{(s)}(l)$ only when
there are branches left going out from a node, in average when $b(s)
\geq 1$. The number of paths with a specific length in the whole
system is therefore
\begin{eqnarray}
r(l) = \sum_{s=0}^{L} r^{(s)}(l) \, n(s),
\label{cumulative_paths}
\end{eqnarray}
where $n(s)$ is defined as earlier in Eq.~(\ref{MF_n_l}), the number
of nodes on a given level $s$.

The Barab\'asi-Albert model allows for more rigorous derivations of
the relation for $n(l)$. The mathematics community often refers to the
tree interpretation of the model as \emph{recursive trees}, and thus
exact results have been obtained for both the distance distribution
and the diameter of the
trees~\cite{pittel_rec_trees,poisson_approx_trees,distr_recursive_trees}.
Bollob\'as and Riordan give a general proof for the diameter scaling
of scale-free BA graphs~\cite{bollobas_riordan}.  The mapping to
Cayley trees also resembles the work of Krapivsky and Redner, who
arrive at a closed recursive analytical form for $n(l)$, in a more
general context than that of scale-free
trees~\cite{organization_growing_nets}. It also resembles Cayley
models of Internet traceroutes~\cite{fractal_internet} by Caldarelli
and co-workers.

\begin{figure}
\begin{minipage}[t]{\linewidth}
\begin{center}
\includegraphics[width=\figwidth]{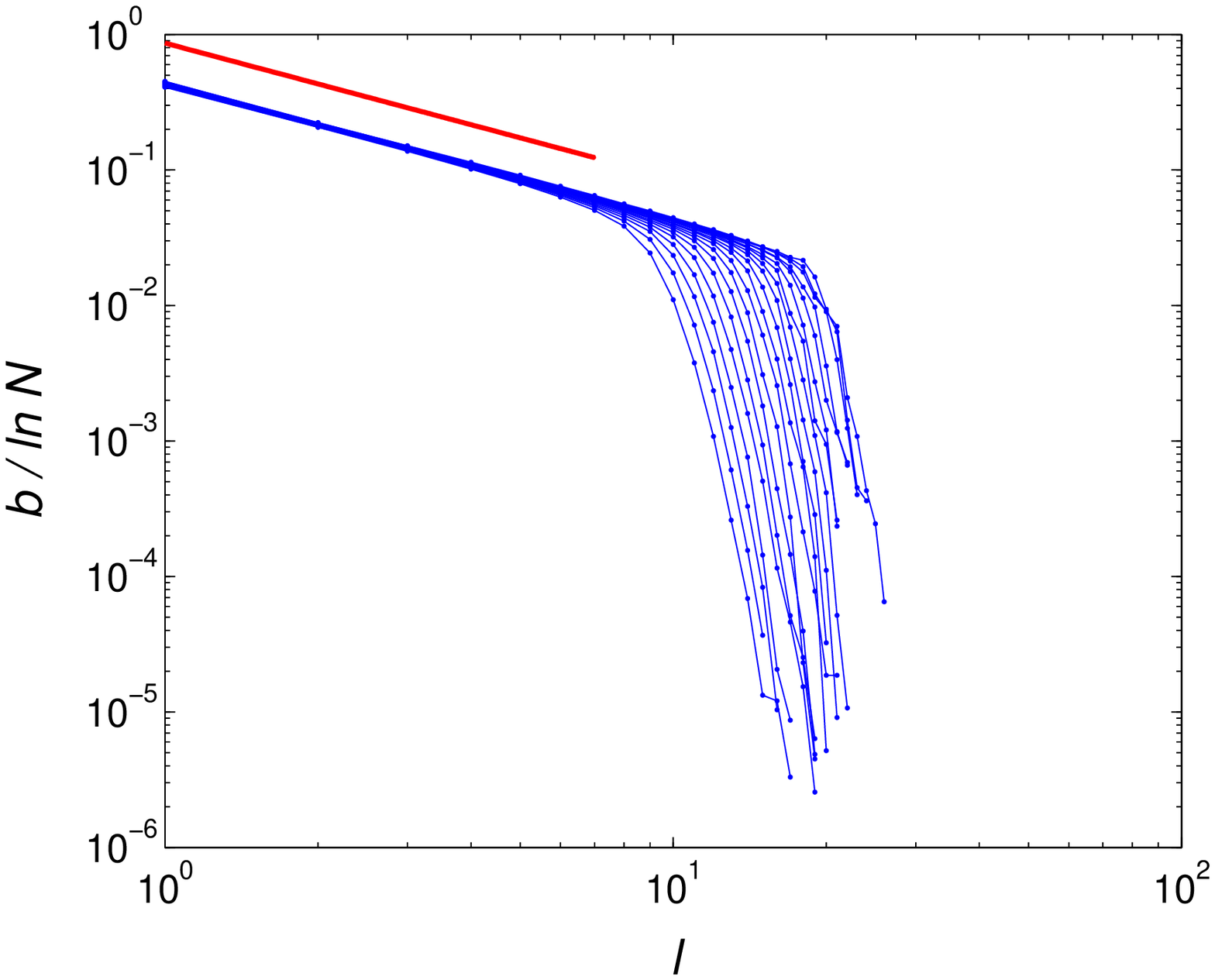}
\end{center}
\vspace{-42mm}
\hspace{-22mm}
\includegraphics[width=0.36\linewidth]{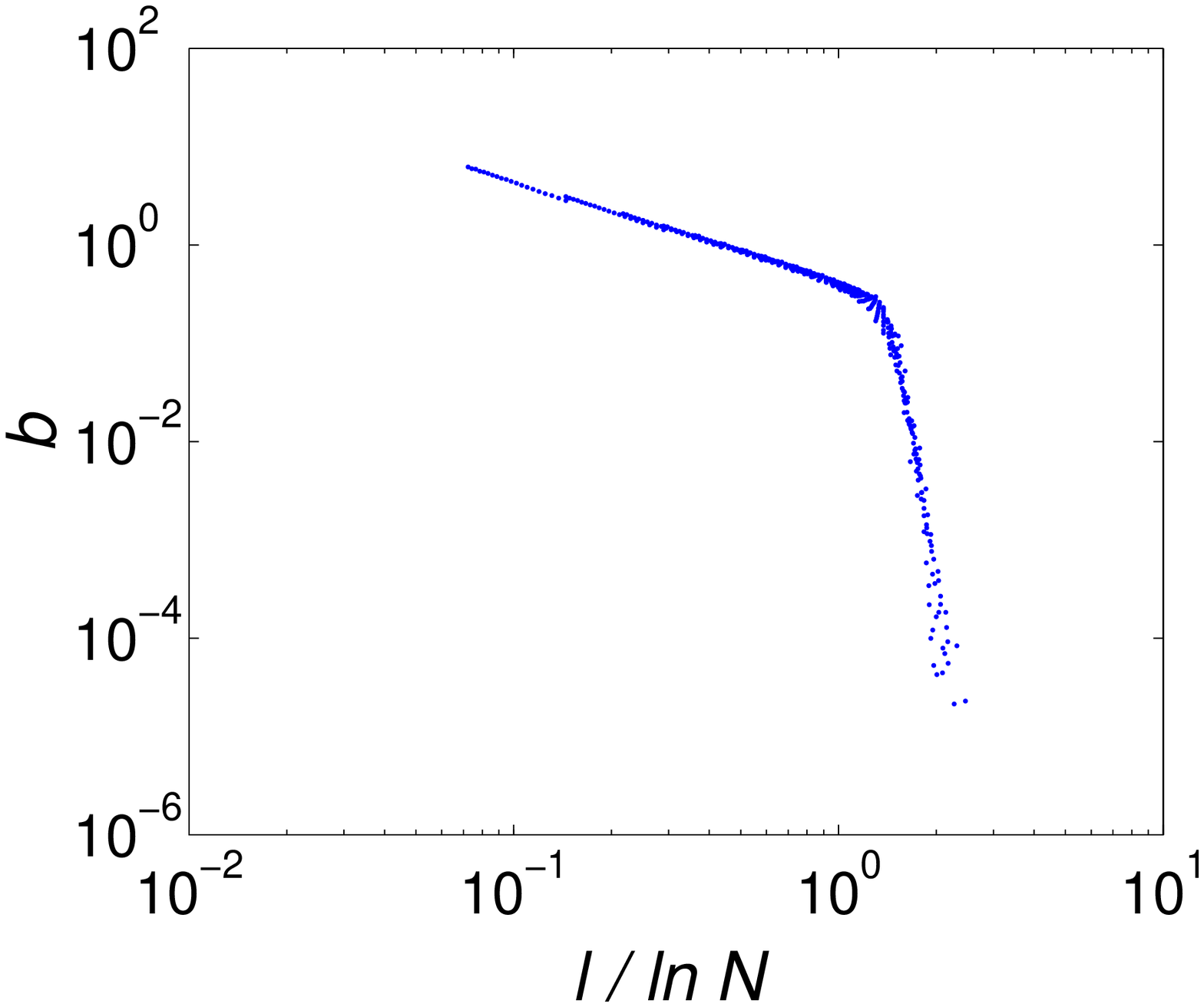}
\end{minipage}
\caption{The average number of branches per node normalized with the
logarithm of the system size, represented versus the minimal distance
of the nodes from the root with maximum connectivity for the BA
model. A power-law fit has been performed in a window indicated by the
heavy line, giving $b(l) / \ln N \simeq 0.43 \, l^{-0.9995}$. The
exponent is very close to $-1$. The inset shows $b(l)$ plotted against
the normalized minimal distance. The systems range from $10^3$ to
$10^6$ nodes in size with logarithmic increments. The number of
iterations for the systems go from $10^5$ to 100, depending on the
size, $N$.}
\label{bratio}
\end{figure}

\section{Comparison with simulations}

Numerical simulations of BA scale-free trees fully confirm the
inferences drawn in the preceding section. Most important of all, the
average number of branches per node on a given level is shown in
Fig.~\ref{bratio}. The numerical parameters of the power-law fit
conform with the mean-field values: the exponent of the decay is
almost exactly $-1$, and the prefactor of the logarithm with $0.43$ is
also close to that of the predicted $1/2$. It is also worth to note
that if we rescale the distance variable by the logarithm of the
system size, we can attain a data collapse with a very good
accuracy. This means that for BA trees in practice $b(l)$ can be
approximated as
\begin{eqnarray}
b(l) = \left\{
\begin{array}{ll}
\frac{0.43 \, \ln N}{l} & \textrm{if $l \lesssim L(N)$}\\
\approx 0               & \textrm{otherwise}
\end{array} \right.
\label{bl_estimate}
\end{eqnarray}

From the inset of Fig.~\ref{bratio} it is also apparent that the
cutoff $L(N)$ is a little over the value of $\ln N$, by a factor of
about $1.3$. On the other hand, the drop of $b(l)$ at $L(N)$ is
measured to be either an exponential or a power-law with a very large
exponent. The mean-field prediction for the maximum of the shortest
path length, $L(N)$, can be obtained by equating $n(L) = 1$ in
Eq.~(\ref{MF_n_l}) and using the Gaussian approximation of
Eq.~(\ref{nl_approx}). The solution up to first order in $\ln N$ is
that $L(N) \approx \frac{1 + \sqrt{2}}{2} \ln N$, which again is in
reasonable agreement with the mean-field argument.

\begin{figure}
\begin{center}
\includegraphics[width=\figwidth]{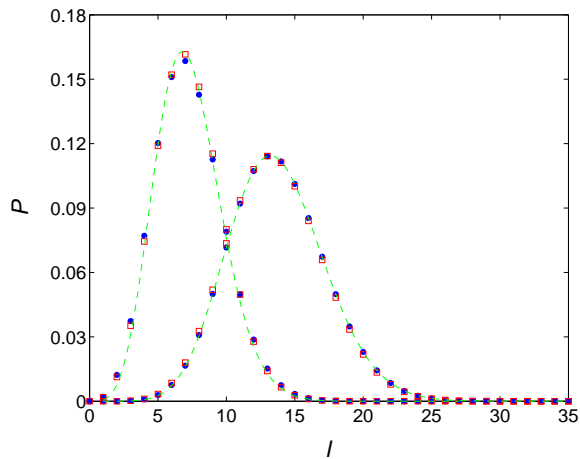}
\end{center}
\caption{Root-to-node (left) and node-to-node (right) distance
distributions with circles (BA model) and their predicted values with
squares. The prediction is based on
Eqs.~(\ref{subtree_leaves})--(\ref{cumulative_paths}). Trees of $10^6$
nodes are measured and averaged over 100 realizations. The dashed
lines show the least-square fits with the function $[C/(l-1)]\,^{l-1}$
to the measured data points. The constant for root-to-node distances
is $C_r=15.7$ and for node-to-node distances $C_n=33.2$. $C_r$ is in a
very good correspondence with the analytical value of $C_r=0.43 \, e
\, \ln N=16.2$ and $C_n \approx 2 C_r$.}
\label{l_Pl}
\end{figure}

The derived quantities $n(l)/N$ and the node-to-node distance
distribution is shown in Fig.~\ref{l_Pl} for two distinct cases:
\begin{itemize}
\item the root-to-node and node-to-node shortest path distribution is
measured in an ensemble of random BA trees using simulations. Instead
of every possible pair, the node-to-node distances are measured only
between a large but finite number of randomly selected vertex pairs,
for practical reasons;
\item both distribution functions are estimated by utilizing the
mean-field tree mapping, using the asymptotic form of
Eq.~(\ref{bl_estimate}) for $b(l)$ with a cutoff at $L = 1.2 \ln N$.
\end{itemize}

It is to be seen that a very good correspondence is found between the
root-node distribution functions, but the overall two-point pdf's are
sensibly close as well.

\begin{figure}
\begin{center}
\includegraphics[width=\figwidth]{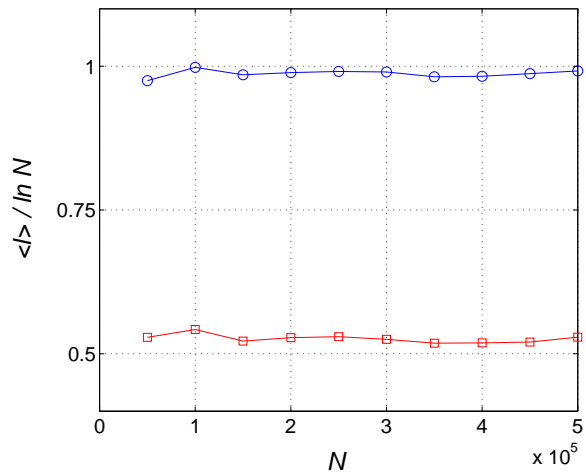}
\end{center}
\caption{The diameter and mean depth of networks of different sizes,
divided by the logarithm of their size. Circles represent the mean of
node-to-node distances, while squares the mean root-to-node distances.
Both are apparently proportional to $\ln N$, and the prefactors of
$\sim 0.5$ and $\sim 1$ are in very good agreement with their
respective analytical values.}
\label{diameter}
\end{figure}

While it has been relatively easy to derive analytical results for the
root-node distances in the mean-field trees,
Eq.~(\ref{cumulative_paths}) and the quantities it is constructed of
turn out to be too complex to handle without numerics. The formulas
(\ref{subtree_leaves})--(\ref{cumulative_paths}) above are used to
calculate the approximate values of the node-to-node path length
distribution in the mean-field trees using the expression of
Eq.~(\ref{bl_estimate}) instead of the analytical form of
Eq.~(\ref{MF_branching}), so as to better represent the random BA
trees. It is reassuring that the generic form of the node-to-node
distance pdf also follows a $(C'/x)^x$ function, only with a different
$C'$ constant from that of $C$ for the root-node distances
[Eq.~(\ref{MF_n_l})--(\ref{nl_approx})]; see Fig.~\ref{l_Pl}. The
diameter of the trees relative to the logarithm of the system size can
be seen on Fig.~\ref{diameter}. $\langle l \rangle \approx \ln N$ or
in other words twice the mean of root-to-node distances. This is
somewhat expected as the main contribution to the node-to-node paths
arises from passing through the root, for large graphs. It leads to a
convolution-type distribution (from the two 'legs'). It can easily be
seen that the diameter cannot exceed twice the depth of the tree,
which gives rise to a logarithmic growth in any case.

\section{On the ``load'' on trees}

On a hierarchical structure the total number of minimum paths going
through a node (the ``load'') can be divided into two contributions.
First, those paths that connect nodes in separate sub-branches of the
node to each other, and, second, those that connect the nodes
belonging to the branches to the rest of the tree. Call $d(l)$ the
number of the descendants of a node on level $l$. In other words,
$d(l)$ is the size of the subtree for the node. Then the load can be
written simply as
\begin{equation}
\Lambda(l) = {b(l) \choose 2} \left[ \frac{d(l)}{b(l)} \right] ^2 +
d(l) \times (N - d(l)),
\label{load_on_tree}
\end{equation}
where the last term counts the connections towards the hub.

For the particular example we are concerned with, it is easy to see
that the latter term dominates ($N \gg d(l)$) and moreover that a good
approximation is given by just simply $\Lambda(l) \approx d(l) \,
N$. Thus one may investigate the dependence of the load on the level
(or depth) of the tree, $l$. For $d(l)$ in the mean-field picture one
has that
\begin{eqnarray}
d(l) = \frac{\sum_{i=l+1}^{L} n(i)}{n(l)},
\label{descendants_for_load}
\end{eqnarray}
and for the layer immediately below
\begin{eqnarray}
d(l + 1) & = & \frac{\sum_{i=l+2}^{L} n(i)}{n(l+1)} =
\frac{d(l) n(l) - n(l+1)}{n(l+1)} = {}
\nonumber\\
& = & \frac{d(l)}{b(l)} - 1 \approx \frac{d(l)}{b(l)},
\end{eqnarray}
where we also used the recursion relation for $n(l+1)$. Finally, the
load changes for the layer underneath as
\begin{eqnarray}
\Lambda(l + 1) = d(l + 1) \, N = \frac{d(l)}{b(l)} N =
\frac{\Lambda(l)}{b(l)}.
\end{eqnarray}
Since the load $\Lambda(l)$ is the same for each of the nodes on a
particular level $l$, the distance--load distribution is directly
given by the normalized $n$--$\Lambda$ function, thus hiding the
implicit dependence on $l$. Considering that
\begin{eqnarray}
\Lambda(l + 1) & = & \frac{\Lambda(l)}{b(l)}
\nonumber\\
n(l + 1) & = & n(l) \, b(l),
\label{load_n_relations}
\end{eqnarray}
$\Lambda(l + 1) \, n(l + 1) = \Lambda(l) \, n(l) = \textit{const.}$
and therefore
\begin{eqnarray}
n = \frac{\textit{const.}}{\Lambda}.
\end{eqnarray}

\begin{figure}
\begin{minipage}[t]{\linewidth}
\begin{center}
\includegraphics[width=\figwidth]{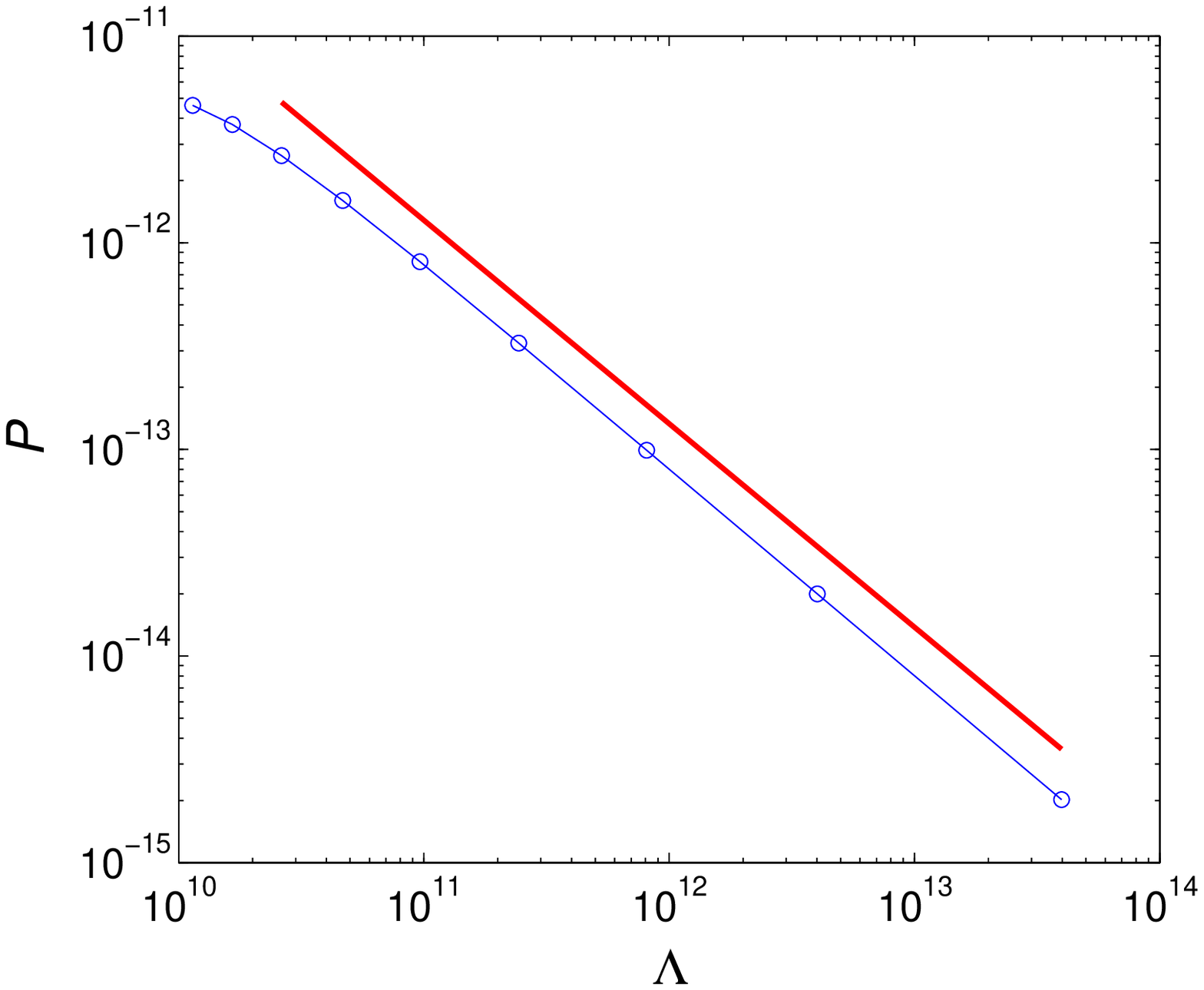}
\end{center}
\vspace{-65mm}
\hspace{39mm}
\includegraphics[width=0.34\linewidth]{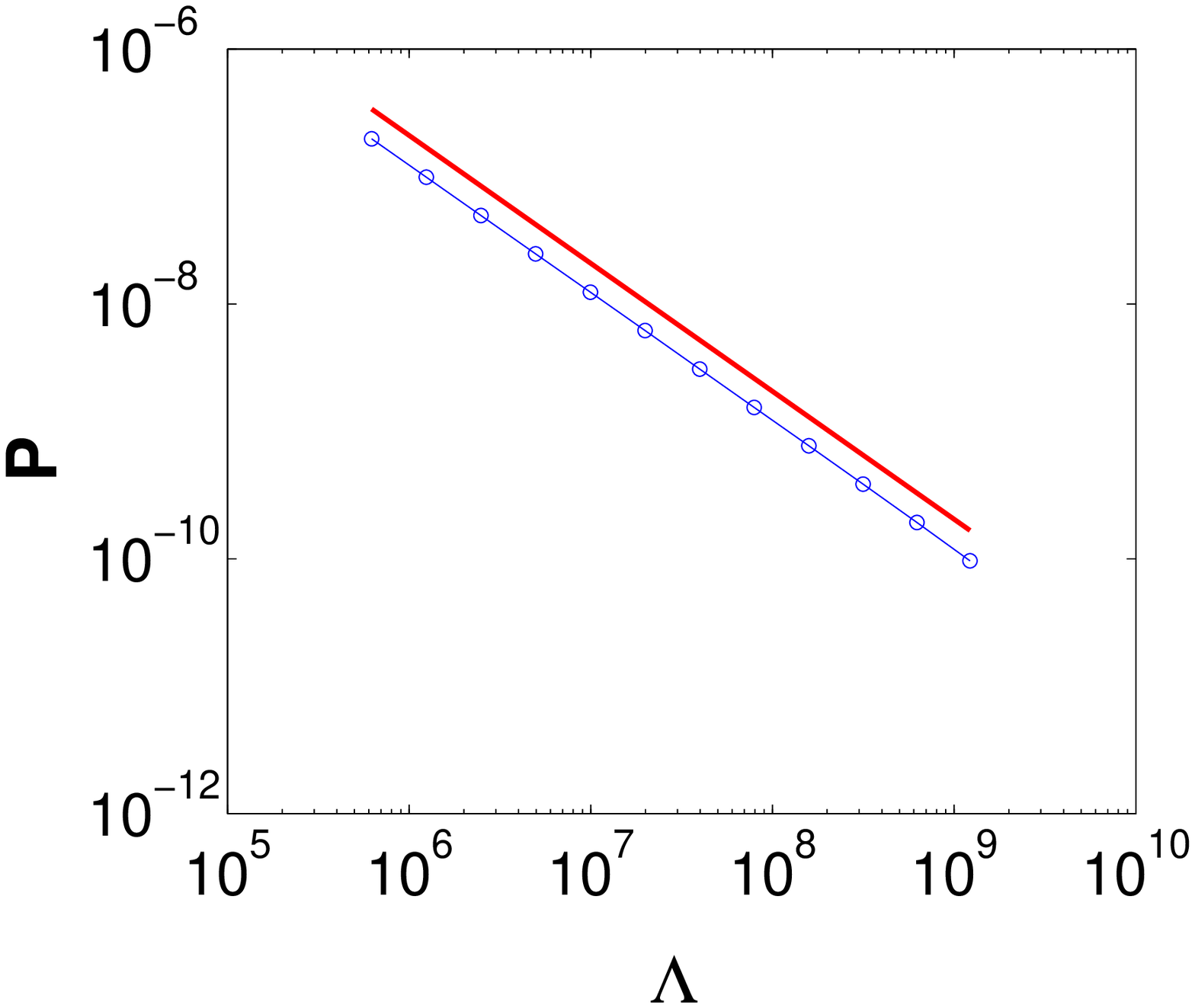}
\end{minipage}
\caption{Load distribution for mean-field trees modeling BA
networks. The probability $P$ is proportional to $n$, the number of
nodes on the levels of the tree. The load on the root has not been
showed since it does not average. The inset shows the load
distribution for a usual Cayley tree with a coordination number
$z=2$. The bold lines indicate power law fits, which give exponents of
$-0.99$ and $-1$ for the BA and the Cayley trees, respectively. The
mean-field tree is a mapping of a random BA tree with $10^{10}$
nodes.}
\label{load}
\end{figure}

We then expect to see that the load is inversely proportional to the
number of nodes on the levels, which is indeed the case according to
Fig.~\ref{load}. The same result holds for normal Cayley trees from
Eq.~(\ref{load_on_tree}).

Note that we have to use mean-field trees which would correspond to
random networks with a large number of nodes so that the number of
levels is of the order of 10. For the load distribution we consider
only levels for which $b(l) \geq 1$ because otherwise subtrees do not
exist in the average sense. It is surprising that the load
distribution exponent does not depend on the actual form of $b(l)$,
being universally $-1$ [Eq.~(\ref{load_n_relations})]. Indeed,
\emph{the exponent of the distance--load pdf is independent of the
choice of the node that all the distances are taken relative to}.

Another common way of defining the importance of the nodes in terms of
shortest paths passing through them is the one called betweenness,
favorable for its algorithmic feasibility and simplicity. Newman
presents a breadth first search algorithm for efficient calculation of
the betweenness of nodes on random graphs~\cite{scientific_coll}. The
only notable difference to Eq.~(\ref{load_on_tree}) comes from the
fact that the betweenness also accounts for paths that originate from
the nodes themselves, which nevertheless amounts only to a constant
system size.

\begin{figure}
\begin{center}
\includegraphics[width=\figwidth]{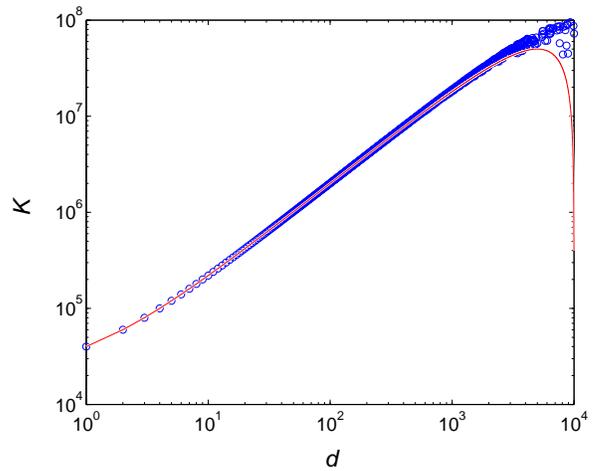}
\end{center}
\caption{Betweenness as a function of the descendants for every node
in the network. Trees of size $10^4$ are taken with 100 realizations.
The root which descendants are defined down from is always the initial
node. The prediction of Eq.~(\ref{betweenness_approx}) is represented
by the solid line.}
\label{betweenness_desc}
\end{figure}

We will calculate the betweenness on the trees, now focusing as a goal
on the probability distribution of the load. An estimation can be
given for a node by considering the contributions to it, and by
separating the network to a descendants part with $d$ nodes in the
branches and all the rest with $N-1-d$ nodes. The node being the
source, we have $N$ shortest paths to any other node; if the source is
among the descendants, we have $d(N-1-d+1)$ ones going through; if the
source is any other node from the network, we have $(N-1-d)(d+1)$. A
fourth contribution, coming from paths passing through the node but
having both ends in the descendant tree, has been neglected. They add
up to an estimated betweenness $K$ of
\begin{eqnarray}
K(d) \approx 2N-1 + 2(N-1)d - 2d^2.
\label{betweenness_approx}
\end{eqnarray}
Here it is to be seen that for small $d$'s the linear term dominates,
just as in our previous load calculation; Fig.~\ref{betweenness_desc}
justifies our estimations.

\begin{figure}
\begin{minipage}[t]{\linewidth}
\begin{center}
\includegraphics[width=\figwidth]{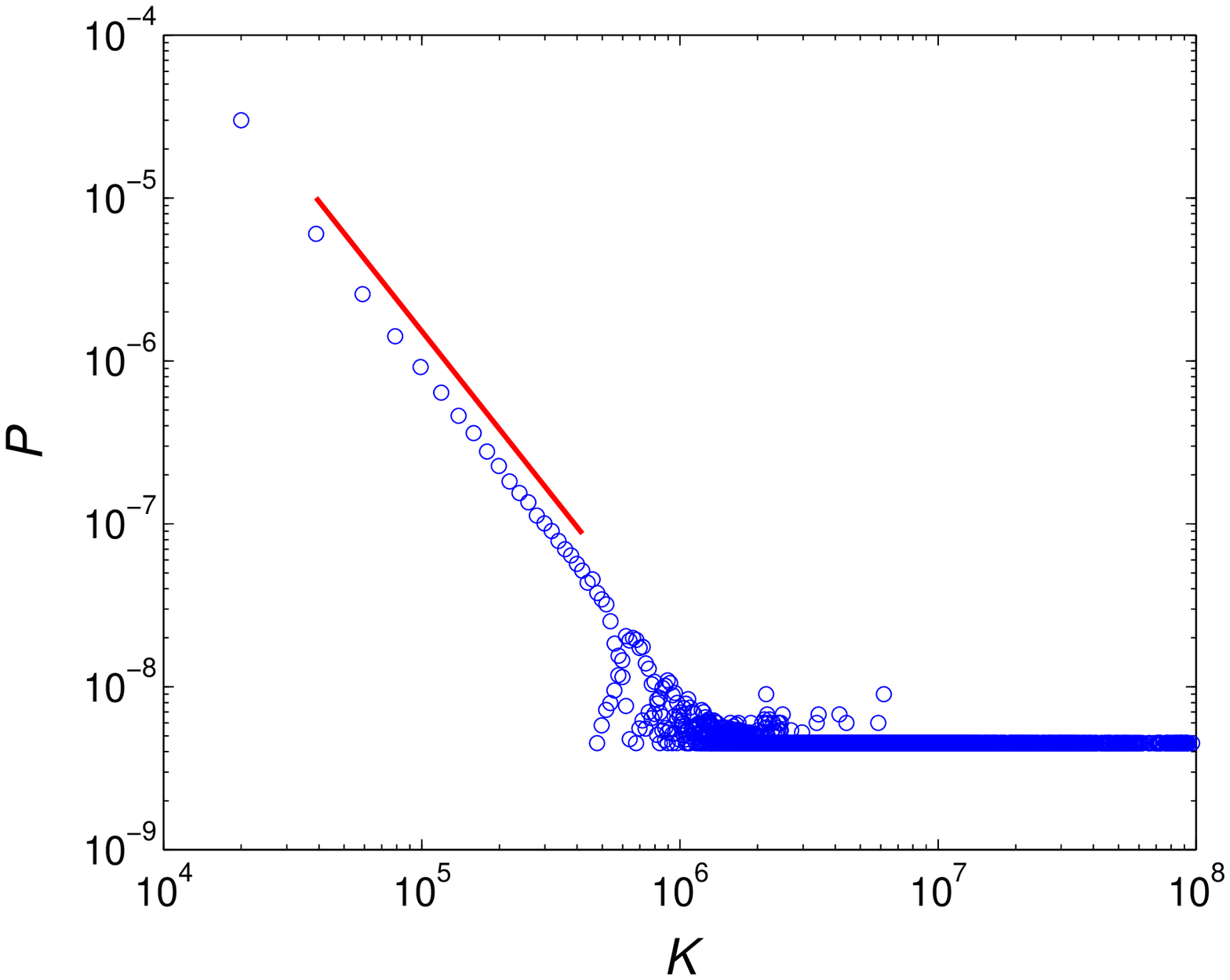}
\end{center}
\vspace{-62mm}
\hspace{32mm}
\includegraphics[width=0.42\linewidth]{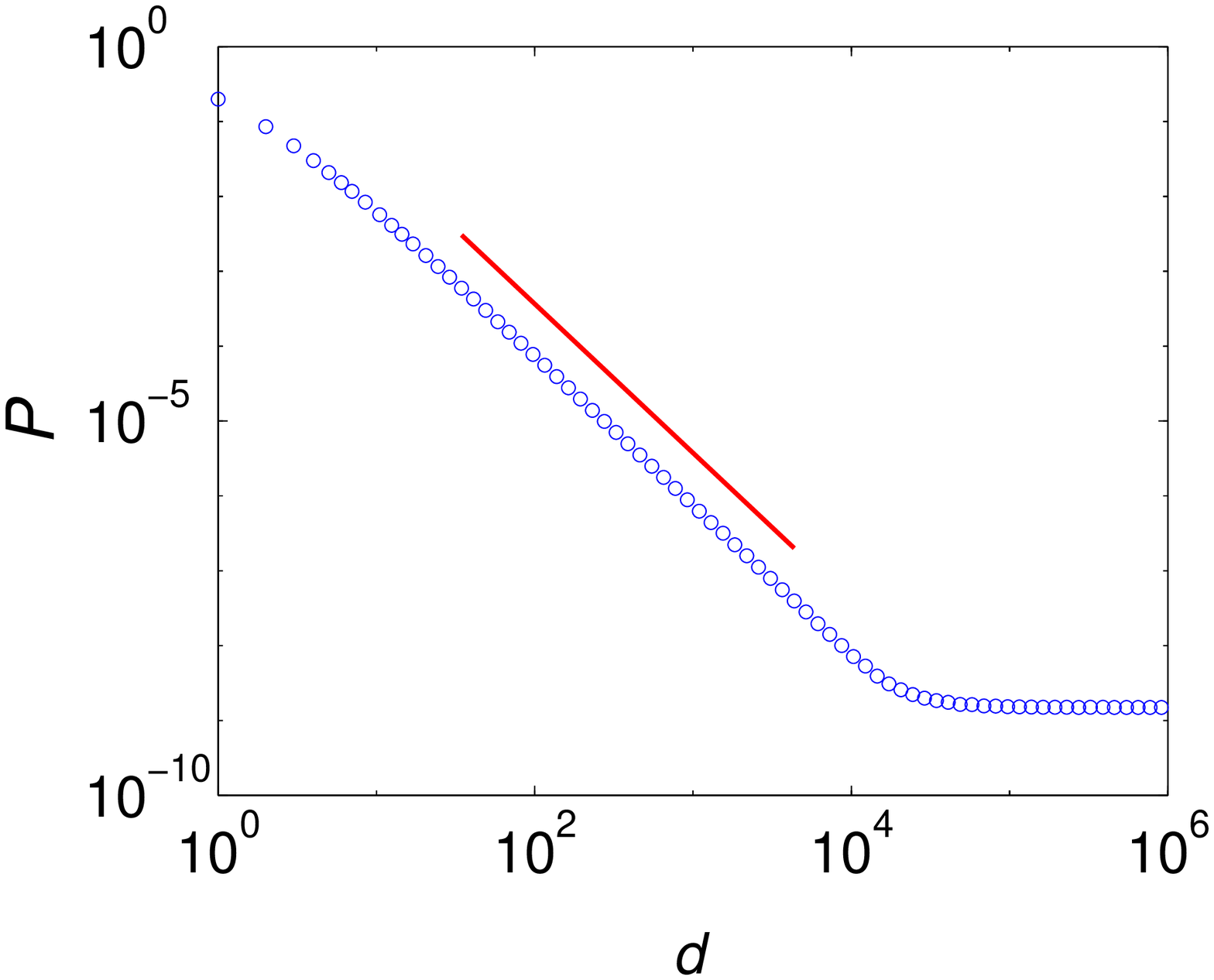}
\end{minipage}
\caption{Betweenness pdf for a system of $10^4$ nodes. The exponent of
the power law shown is $-1.99$. The inset displays the logarithmically
binned pdf of the descendants for systems with $10^6$ nodes. Its power
law exponent is $-1.99$ as well.}
\label{betweenness_distrib}
\end{figure}

The betweenness probability distribution $P(K)$ taken over all nodes
in the network can then be concluded to asymptotically follow a power
law decay with a universal exponent of $-2$. This is since $K$ is
linear in the number of descendants $d$ and moreover that the pdf of
$d$ scales universally with an exponent of $-2$ for supercritical
trees~\cite{supercritical_PL}. Strictly speaking, the conclusions here
are only true for the supercritical part of the tree, i.e. where $b(l)
> 1$. The subcritical leaves of the tree have an increasingly smaller
number of descendants, though, which drop exponentially with each new
layer, and it can be verified that the descendant pdf decay exponent
is indeed above 2 if only this part of the tree is considered.
Nevertheless, Fig.~\ref{betweenness_distrib} shows that both the
descendant pdf and the load pdf are accurately described by inverse
square functions. A scaling of the load distribution has been
experimentally found on other scale-free networks as
well~\cite{universal_load}, only with a slightly different universal
exponent of about $2.2$.

\begin{figure}
\begin{minipage}[t]{\linewidth}
\begin{center}
\includegraphics[width=\figwidth]{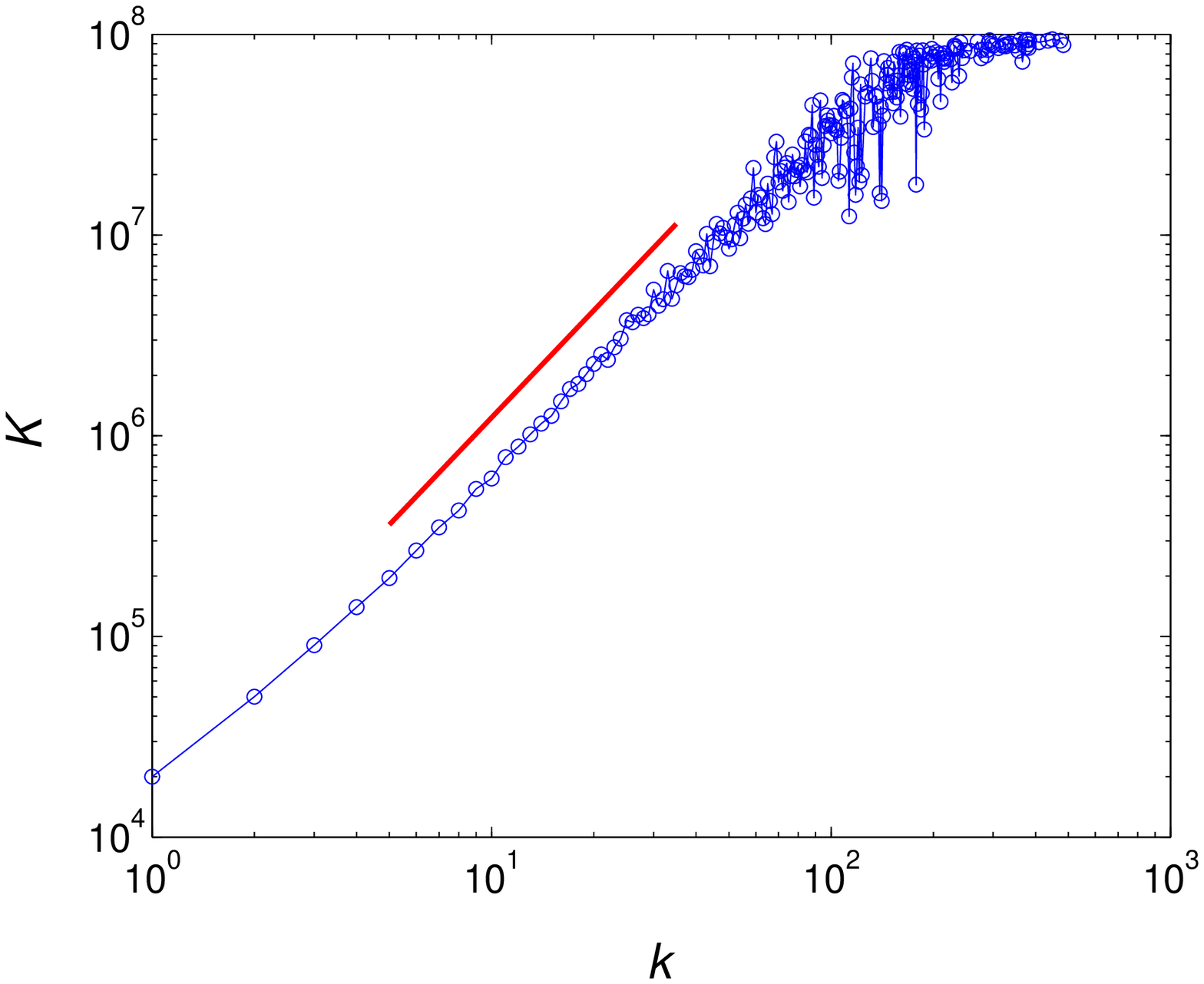}
\end{center}
\vspace{-44mm}
\hspace{34mm}
\includegraphics[width=0.40\linewidth]{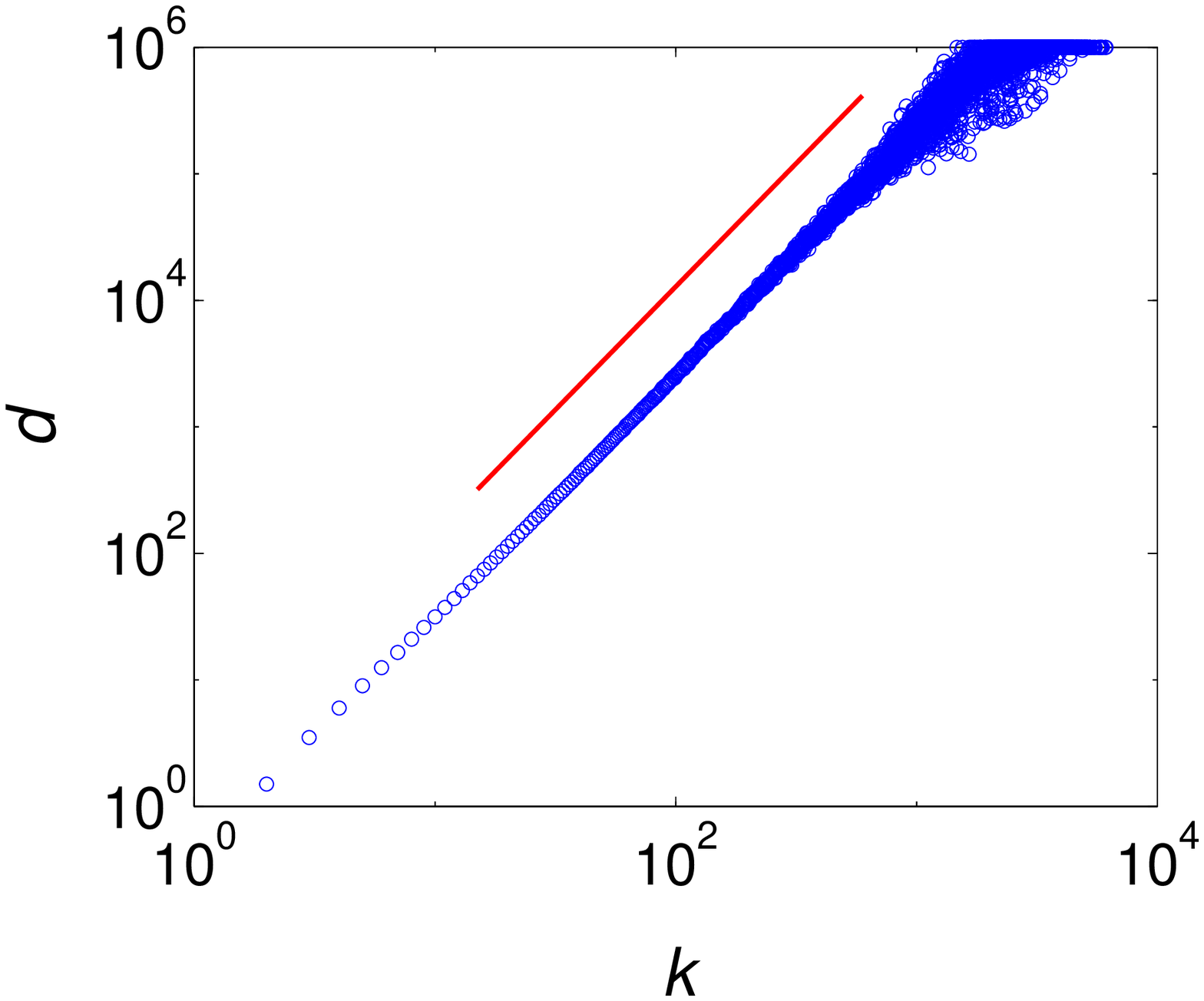}
\end{minipage}
\caption{Average betweenness for nodes with a particular degree, taken
over 100 realizations of networks with $10^4$ nodes. The fit of a
power law indicates an exponent of about $1.78$. The inset shows the
average number of descendants versus the degree for systems with
$10^6$ nodes, for which a power-law fit gives $d(k) \sim k^{1.95}$.}
\label{betweenness_func}
\end{figure}

A further, practically more far-reaching observation is that the
average betweenness as measured as a function of the locally known
node degree grows as a power-law of the degree with an exponent of
about $1.8$ (Fig.~\ref{betweenness_func}). A mean-field approach can
be used to estimate for the exponent, though, if we consider that the
preferential attachment principle for large degrees gives rise to a
descendant--degree scaling of $d \sim k ^ \frac{1}{\beta}$
($\beta=1/2$), which is the inverted relation for the time evolution
of the degree of a parent node~\cite{stat_mech}. In this particular
case, time is measured as the size of the node's subtree. A
substitution of the latter into the linear load equation would suggest
an exponent of $2$; the deviation from it may come from the rather
restricted range of the degree that the relatively small system sizes
allow.

\section{Conclusions}

In this paper we have mapped scale-free Barab\'asi-Albert trees to a
deterministic model of a rooted tree with a uniform branching process
on each layer of the tree. This idea resembles work on the Internet
structure~\cite{fractal_internet} and on the structure of branched
cracks, where an inverse relation of the branching to distance has
been observed~\cite{bouchaud}.

Simulations show that the distribution of the number of branches on
one particular layer of the tree follows a power-law function, but it
turns out to be a good approximation to describe the branching only by
its mean, $b(l)$. In the simple case of BA trees it can be shown by
means of this mapping that the diameter of the networks is bounded by
the logarithm of the network size and the asymptotic form of the
distance distribution functions follows immediately. In other words,
we can examine the slow convergence of this function to the limiting
Gaussian form for infinite system sizes. Given an effective
description in terms of a tree plus a branching process, further
information can be found, e.g. one may consider the scaling of the
number of shortest-distance paths (load, or betweenness).  Non-uniform
critical trees could perhaps be constructed in a self-organized
fashion, as is possible for the statistically uniform case
~\cite{SO_branching}.

One should note the close relation of the Cayley representation to
minimal spanning trees (MST) on scale-free (random) networks; for the
$m=1$-networks these two coincide. This makes it an interesting
prospect to study the load and distance properties of MST's in other
scale-free networks.
\bigskip

We thank B\'alint T\'oth for useful discussions.

This work has been supported by the Academy of Finland's Centre of
Excellence Programme and by OTKA T029985.



\begin{thebibliography}{99}

\bibitem{tunable_clustering} P.\ Holme and B.J.\ Kim,
cond-mat/0110452.

\bibitem{clustered_scale_free} K.\ Klemm and V.M.\ Egu\'\i luz,
Phys. Rev. E {\bf 65}, 036123 (2002); cond-mat/0107607.

\bibitem{develop_decay} S.N.\ Dorogovtsev and J.F.F.\ Mendes,
Europhys. Lett. {\bf 52}, 33 (2000).

\bibitem{heritable_connectivity} S.N.\ Dorogovtsev, J.F.F.\ Mendes,
and A.N.\ Samukhin, cond-mat/0011077.

\bibitem{Krzywicki1} Z.\ Burda, J.D.\ Correia, and A.\ Krzywicki,
Phys. Rev. E {\bf 64}, 046118 (2001).

\bibitem{generic_scale} S.N.\ Dorogovtsev, J.F.F.\ Mendes, and A.N.\
Samukhin, cond-mat/0011115; Phys. Rev. E {\bf 63}, 062101 (2001).

\bibitem{scaling_evolving} S.N.\ Dorogovtsev and J.F.F.\ Mendes,
Phys. Rev. E {\bf 63}, 056125 (2001).

\bibitem{internet} A. Vazquez, R.\ Pastor-Satorras, and A. Vespignani,
cond-mat/0112400.

\bibitem{stat_mech} R.\ Albert and A.-L.\ Barab\'asi,
cond-mat/0106096; Rev. Mod. Phys. {\bf 74}, 47 (2002).

\bibitem{evolution} S.N.\ Dorogovtsev and J.F.F.\ Mendes,
cond-mat/0106144.

\bibitem{BA_original} A.-L.\ Barab\'asi and R.\ Albert, Science {\bf
286}, 509 (1999).

\bibitem{Krzywicki2} A.\ Krzywicki, cond-mat/0110574 (2001).

\bibitem{epidemic_spreading} R.\ Pastor-Satorras and A. Vespignani,
Phys. Rev. Lett. {\bf 86}, 3200 (2001).

\bibitem{epidemic_dynamics} R.\ Pastor-Satorras and A. Vespignani,
cond-mat/0202298.

\bibitem{community_structure} M.\ Girvan and M.E.J.\ Newman,
cond-mat/0112110.

\bibitem{local_events} R.\ Albert and A.-L.\ Barab\'asi,
Phys. Rev. Lett. {\bf 85}, 5234 (2000).

\bibitem{path_finding} B.J.\ Kim, C.N.\ Yoon, S.K.\ Han, and H.\
Jeong, cond-mat/0111232.

\bibitem{exact_small_world} R.V.\ Kulkarni, E.\ Almaas, and D.\
Stroud, Phys. Rev. E {\bf 61}, 4268 (2000).

\bibitem{pseudofractal} S.N.\ Dorogovtsev, A.V. Goltsev, and A.N.\
Samukhin, cond-mat/0112143.

\bibitem{universal_load} K.-I.\ Goh, B.\ Kahng, and D.\ Kim,
cond-mat/0106565; Phys. Rev. Lett. {\bf 87}, 278701 (2001).

\bibitem{scientific_coll} M.E.J.\ Newman, Phys. Rev. E {\bf 64},
016132 (2001).

\bibitem{arbitrary_degree} M.E.J.\ Newman, S.H.\ Strogatz, and D.J.\
Watts, Phys. Rev. E {\bf 64}, 026118 (2001).

\bibitem{BRV} A.L. Barab\'asi, E. Ravasz, and T. Vicsek,
cond-mat/0107419.

\bibitem{pittel_rec_trees} B. Pittel, Random Struct. Algorithms {\bf
5}, 337 (1994).

\bibitem{poisson_approx_trees} R.P. Dobrow and R.T. Smythe, Random
Struct. Algorithms {\bf 9}, 79 (1996).

\bibitem{distr_recursive_trees} R.P. Dobrow, J. Appl. Probab. {\bf
33}, 749 (1996).

\bibitem{bollobas_riordan} B. Bollob\'as and O. Riordan, to be
published.

\bibitem{organization_growing_nets} P.L. Krapivsky and S. Redner,
Phys. Rev. E {\bf 63}, 066123 (2001).

\bibitem{fractal_internet} G. Caldarelli, R. Marchetti, and
L. Pietronero, Europhys. Lett. {\bf 52}, 386 (2000).

\bibitem{supercritical_PL} P. De Los Rios, Europhys. Lett. {\bf 56},
898 (2001).

\bibitem{bouchaud} J.P. Bouchaud, E. Bouchaud, G. Lapasset, and
J. Plan\`es, Phys. Rev. Lett. {\bf 71}, 2240 (1993).

\bibitem{SO_branching} S. Zapperi, K.B. Lauritsen, and H.E. Stanley,
Phys. Rev. Lett. {\bf 75}, 4071 (1995).

\end{thebibliography}
\end{document}